\documentclass{PoS}
\usepackage{amsmath,amssymb}  
\usepackage{bm,bbm}  
\usepackage{slashed}
\newcommand\beq{\begin{equation}}
\newcommand\eeq{\end{equation}}

\newcommand\diag{\text{diag}}
\newcommand\hc{\text{h.c.}}

\newcommand\PD{{\phantom{\dagger}}}
\newcommand\sfrac[2]{{\textstyle{\frac{#1}{#2}}}}

\title{Hidden Fine Tuning In The Quark Sector Of Little Higgs Models}

\ShortTitle{Hidden Fine Tuning In Little Higgs Models}

\author{\speaker{Benjam\'\i{}n Grinstein}\thanks{The author wishes to
    thank the theory group at CERN for their hospitality. Work
    supported in part by
    U.S. Department of Energy under Contract No. DE-FG03-97ER40546. }\\
  University of California, San Diego and CERN\\
  E-mail: \email{bgrinstein@ucsd.edu}}

\author{Randall Kelley\\
        Harvard University\\
        E-mail: \email{randallkelley@physics.ucsd.edu}}

\author{Patipan Uttayarat\\
        University of California, San Diego\\
        E-mail: \email{puttayarat@physics.ucsd.edu}}

      \abstract{In Little Higgs models a collective symmetry prevents
        the higgs from acquiring a quadratically divergent mass at one
        loop. We have previously shown that the couplings in the
        Littlest Higgs model introduced to give the top quark a mass
        do not naturally respect the collective symmetry. We extend
        our previous work showing that the problem is generic: it
        arises from the fact that the would be collective symmetry of
        any one top quark mass term is broken by gauge interactions.
      }

\FullConference{35th International Conference of High Energy Physics - ICHEP2010,\\
		July 22-28, 2010\\
		Paris France}

\begin{document}
\section{Fine Tuning Problem in The Littlest Higgs Model}
To establish notation we briefly review elements of the Littlest Higgs
\cite{ArkaniHamed:2002qy}. It has a global symmetry $G_f=SU(5)$ that
spontaneously breaks to $SO(5)$; $SU(5)\to SO(5)$ is characterized by
the Goldstone boson decay constant $f$.  The embedding of the weakly
gauged subgroup $G_w=\prod_{i=1,2}SU(2)_i\times U(1)_i$ in $G_f$ is
fixed by taking the generators of $SU(2)_1$ and $SU(2)_2$ to be
\begin{equation}
   Q^a_1=  
  \begin{pmatrix}
    \frac{1}{2}\sigma^a&0_{2\times3}\\0_{3\times2}&0_{3\times3}
  \end{pmatrix}
  \qquad\text{and}\qquad
  Q^a_2=
  \begin{pmatrix}
    0_{3\times3}&0_{3\times2}\\0_{2\times3}&-\frac{1}{2}\sigma^{a*}
  \end{pmatrix}\,,
\end{equation}
and the generators of $U(1)_1$ and $U(1)_2$ to be
\begin{equation}
  Y_1=\frac1{10}\diag(3,3,-2,-2,-2)\qquad\text{and}\qquad
  Y_2=\frac1{10}\diag(2,2,2,-3,-3)\,. 
\end{equation}

The vacuum manifold is characterized by a unitary, symmetric
$5\times5$ matrix $\Sigma$. We denote by $g_{i}$ ($g'_{i}$) the gauge
couplings associated with $SU(2)_i$ ($U(1)_i$). If one sets
$g_{1}=g_1'=0$ the model has an exact global $SU(3)_u$ symmetry (acting
on the upper $3 \times 3$ block of $\Sigma$), while for $g_{2}=g_2'=0$ it
has a different exact global $SU(3)_d$ symmetry (acting on the lower $3
\times 3$ block). 
Either of these exact global $SU(3)$ would-be
symmetries guarantee the Higgs remains exactly massless. Hence, the
Higgs mass should vanish for either $g_{1}=g_1'=0$ or
$g_{2}=g_2'=0$. The perturbative quadratically divergent correction to
the Higgs mass must be polynomial in the couplings and can involve
only one of the couplings at one loop order. Hence it must vanish at
one loop. This is the collective symmetry mechanism that ensures the
absence of 1-loop quadratic divergences in the higgs mass.

Next introduce couplings of $\Sigma$ to quarks, to generate a top
mass.  Take the third generation doublet $q_L$ to be a doublet under
$SU(2)_1$ and a singlet under $SU(2)_2$.  Introduce additional
$SU(2)_1\times SU(2)_2$-singlet spinor fields: $q_R$, $u_L$ and
$u_R$. The third generation right handed singlet is a linear
combination of $u_R$ and $q_R$.  The charges of $q_L$, $q_R$, 
$u_L$ and $u_R$ under $U(1)_1\times U(1)_2$ are, in terms of a free
parameter $y$, $(\frac{11}{30}-y,y-\frac15)$, $(\sfrac23-y ,y )$,
 $(\sfrac{13}{15}-y ,y-\sfrac15 )$
and $(\sfrac{13}{15}-y ,y-\sfrac15 )$, respectively. If their
couplings are taken to be
\begin{equation}
  \label{eq:Ltop-split}
  {\cal L}_{\text{top}}=-\lambda_1f\bar q_L^{\PD i}\epsilon^{xy}
  \Sigma_{ix}\Sigma_{3y}q_R^\PD
  -\frac12\lambda'_1f\bar u_{L}^\PD\epsilon^{3jk}\epsilon^{xy}
  \Sigma_{jx}\Sigma_{ky}q_R^\PD
  -\lambda_2f\bar u_L^\PD u_R^\PD +\hc
\end{equation}
then only when $\lambda_1'=\lambda_1$ (and $\lambda_2=g_1=g_1'=0$) do
we obtain the global $SU(3)_u$ symmetry of the collective symmetry
mechanism. In Ref.~ \cite{Grinstein:2009ex} we pointed out that the
relation $\lambda_1'=\lambda_1$, assumed throughout the little higgs
literature, is unnatural. Indeed, if $\Lambda$ is the cutoff of the
theory, then the renormalization group gives
\begin{equation}
\label{eq:su3bkg}
\frac{\lambda_1(\mu)}{\lambda'_1(\mu)}
       =\frac{\lambda_1(\Lambda)}{\lambda'_1(\Lambda)}
\left(\frac{g'_1(\mu)}{g'_1(\Lambda)}\right)^{(2-3y)/b}\,,
\end{equation}
where $b$ is the one-loop coefficient of the $\beta$-function of
$g'_1$. Moreover, this running must occur in the UV completion as
well, and there are additional corrections from matching at
$\Lambda$. So there is no natural way of justifying
$\lambda_1'(\Lambda)=\lambda_1(\Lambda)$.  We refer to this as the
hidden fine tuning problem. How bad is it? There is now a
quadratically divergent contribution to the higgs mass, $ \delta m_h^2
=\frac{12}{16\pi^2}(\lambda^2_1-\lambda^{\prime2}_1)\Lambda^2$. This
requires a tuning $\delta\lambda_1\approx
\frac1{24}\frac{m_h^2}{f^2}\sim0.04\%$ for $m_h=114$~GeV, or a
$\Delta=2400$ naturalness measure\cite{Barbieri:1987fn}.  Running is a
1-loop effect and $\lambda_1-\lambda_1'$ contributes to mass at
1-loop. Enthusiasts of the model may argue that therefore the actual
correction to the higgs mass is a 2-loop effect.  But in the absence
of fine tuning at $\Lambda$ it is really a 1-loop effect.  Moreover,
numerically the effect is large: $\delta \lambda_1< 4\times 10^{-4}$
is needed, while 1-loop is $1/16\pi^2 \approx63\times 10^{-4}$. Note
also, from Eq.~\eqref{eq:su3bkg}, that while $y=2/3$ gives no 1-loop
logarithmic running, one cannot ignore finite, non-logarithmic
corrections. We computed the logarithmic corrections because they are
universal. But there is no reason to expect that the running above
$\Lambda$ plus the matching at $\Lambda$ will keep
$\lambda_1'=\lambda_1$ even when the special value $y=\frac23$ is chosen.

Can one impose a symmetry in the underlying UV theory that enforces
$\lambda_1'=\lambda_1$ to high accuracy in spite of the fact that the
symmetry is broken by gauge interactions?  Let us look at a more
familiar example.  Consider $SU(3)$ as an approximate flavor symmetry
of QCD.  This is a natural symmetry, in the sense that it appears
automatically because all quarks are light compared to the chiral
symmetry breaking scale, regardless of the relative magnitude of the
masses. In the absence of fine tuning, flavor-symmetry breaking
interactions in a phenomenological Lagrangian take the most general
form consistent with gauge invariance. Short of an accidental tuning the only
alternative is  a perturbative UV completion. This however
involves fundamental scalars that Little Higgs theories set out to
avoid.

\section{A No-go theorem}
Consider a model with global ``flavor'' symmetry group $G_f$. It is
assumed to break spontaneously, $G_f\to H$. There is a weakly gauged
subgroup $G_w\subset G_f$, and $G_w\to G_{\text{EW}}$ under $G_f\to H$
(where $ G_{\text{EW}}$ stands for the SM's electroweak group). We
assume further that among the pseudo-goldstone bosons in $G/H$ there
is a higgs doublet, $h$. 

In general the gauge group has a product structure,  $G_w=\prod
G_i$. For each $G_i$ we assume there is a collective symmetry group,
$G_i^c$, that commutes with $G_i$ and that induces non-linear shifts
in $h$. This assumption requires that each of the $G_i$ has (four) generators
that are not orthogonal to the generators  of $G_{\text{EW}}$ (other
gauged factors that have no direction along the electroweak group are
of no interest here). 

The theorem is concerned with the possibility of writing additional
terms in the Lagrangian, like the terms required to give the top quark
a mass. If the higgs mass is to be protected from these, then they
each must have their own collective symmetry group $G^c_Y$, and they
must remain invariant under $\prod_{i=1}^N G_i$.  We will show that
one cannot find a $G^c_Y$ that commutes with $\prod_i
G_i$.\footnote{In Ref.~\cite{Grinstein:2009ex} we argued that the
  generators of $G^c_Y$ form a reducible representation of
  $G_{\text{EW}}$, but we did not make explicit the role played by the
  $G_i$ and $G_i^c$. The argument presented in this talk does. It thus
  allows us to elucidate how models like that of Kaplan and
  Schmaltz\cite{Kaplan:2003uc} evade the no-go theorem; see
  Sec.~\ref{sec:KS}.}  Hence a $G^c_Y$ invariant is a sum over terms
related by $G^c_Y$ that are independently gauge
invariant. Alternatively, one could absorb in $G^c_Y$ that part
of the gauge group that does not commute with $G^c_Y$,  gauging
$G^c_Y$, which results in eating the higgs doublet.
 
\paragraph{Proof.} 
That the higgs transforms linearly under the electroweak
gauge group means that under $SU(2)\times U(1)$ the doublet $h$ transforms
as
\begin{equation}
\delta_\epsilon h= i\epsilon^a\frac{\sigma^a}2h+i\epsilon\frac12h\,,
\end{equation}
where $\sigma^a$ are Pauli matrices. Under any of the collective groups
$G^c_i\subset G_f$, $h$ transforms non-linearly,
\begin{equation}
\delta_\eta h= \eta^mx^m+\cdots
\end{equation}
where the implicit sum over $m$ is over all generators in $G^c_i$, for
some two component complex vectors $x^m$ and the ellipses stand for
terms at least linear in $h$.  One can redefine the basis of
generators in $G^c_i$ so that $x^m=0$ for $m\ge5$ and $x^m$ for
$m=1,\cdots, 4$ are unit vectors, with $m=1,3$ real and $m=2,4$
purely imaginary. Now  the commutator,
\begin{equation}
(\delta_\eta\delta_\epsilon-\delta_\epsilon\delta_\eta)h= i\epsilon^a
\eta^m\frac{\sigma^a}2x^m+i\epsilon\eta^m\frac12x^m+\cdots\,,
\end{equation}
is again a non-linear transformation, a linear combination of the same
four generators in $G^c_i$ that shift the higgs. In terms of the Lie
algebra of $G_f$, denoting these generators by $X^i$,
with\footnote{The index $i$ runs over 1,2 because the hermitian
  matrices break into a symmetric and an antisymmetric part,
  corresponding to the two real and two imaginary components of $x^m$,
  and also to the real and imaginary components of the higgs doublet.}
$i=1,2$ and the generators of $G_{\text{EW}}$ by $Q^a$ and $Y$, we
read off
\begin{equation}
\label{comm}
[Q^a,X^i]=\frac{i}2(\sigma^a)^{ij}X^j,\qquad [Y,X^i]=\frac{i}2X^i
\end{equation}
We see that the $G^c_i$-generators of higgs shifts transform as tensors
of $G_{\text{EW}}$ with the same quantum numbers as the higgs doublet.

Let's introduce some more notation. The generators of $G_f$, $H$ and
$G_i$ are denoted by $\{T^A,X^B\}$, $\{T^A\}$ and $\{Q_i^I\}$,
respectively.  $G_{\text{EW}}$ has generators
$\{T^1,\ldots,T^4\}=\{Q^a,Y\}$. We can always arrange the broken
generators $X^B$ so that the first four precisely correspond to the
generators of the non-linear transformations on the higgs doublet,
$X^x$, $x=1,\ldots,4$.\footnote{These four generators are given in
  terms of those in Eq.~\ref{comm} by $X^{1,2}\pm (X^{1,2})^T$.} The
$X^x$ are not necessarily in the algebra of $G_i^c$ but there are some
unbroken generators for which $X^x+T^x$ are. The only gauged
sub-groups that are relevant to our arguments are those that have a
component of $G_{\text{EW}}$. Hence, $Q^a = \sum_{i,I}c^{aI}_i
Q^I_i$. For each $i$ a similarity transformation brings this to
the form $\sum_I c^{aI}_iQ^I_i= c_iQ^a_i$
(no sum on $i$) so that now $ Q^a=\sum_{i}c_i Q^a_i$ with
all $c_i\ne0$. Since $[Q^a_i,Q^b_j]=0$ for $i\ne j$ it follows that
$[Q^a_i,Q^b_j]=i\delta_{ij}\epsilon^{abc}Q^c_i$ and $c_i=1$ (we are
assuming a common normalization for generators). 

Is there a ``yukawa'' collective symmetry group $G^c_Y$ that commutes
with all the gauge groups? The answer is that there is none since a
collective symmetry has to include $X^x$ and therefore the algebra is
going to include that of some (possibly all) of the gauged symmetries.
The proof is straightforward. The generator in $G^c_Y$ that shifts the
higgs, $X^x_Y$, must satisfy \(
[Q^a,X^x_Y]={\textstyle\frac{i}2}(\sigma^a)^{xy}X^y_Y \) Using
$Q^a=\sum Q^a_i$ we see that this is inconsistent with
$[Q^a_i,X^x_Y]=0$. End proof.\footnote{Alternatively, write an $SU(2)$
  subalgebra $X^a_Y$ of $G^c_Y$in terms of the generators of the
  $G^c_i$'s:
  \( X^a_Y= \sum_i c_i X^a_i \).
  Requiring that this commutes with $Q^b_j$ for every $j$ one find
  $c_i=0$, all $i$.  This is not a complete proof inasmuch as we have
  assumed, not proved, that one can find an $SU(2)$ subalgebra of
  $G^c_Y$ that is a linear combination of generators of the
  $G^c_i$.   }

\paragraph{Comments.}
What does this mean? If $[G^c_Y,G_i]\ne0$ then a non-trivial invariant
of $G^c_Y$ is a sum of several terms independently invariant under
$G_i$.  To see this note that, since there is no semi-simple Lie
algebra of rank 4, there must be additional generators in the algebra
that contains $X^x$. Using the Jacobi identity we see that $\hat
X^{xy}=[X^x,X^y]$ satisfies $[Q_i^a,\hat
X^{xy}]=\frac{i}2(\sigma^a)^{yz}\hat
X^{xz}-\frac{i}2(\sigma^a)^{xz}\hat X^{yz}$. That is, these generators
transform in a representation in the tensor product of two
doublets. Continuing this way, considering commutators of the
generators we have so far, we can eventually generate the complete Lie
algebra and find that it breaks into sectors classified by irreducible
representations under $G_i$. Now, any non-trivial invariant must be a
product of two (combinations of) fields, one transforming in some
irreducible representation $R$ of $G^c_Y$ and the other as the complex
conjugate $\bar R$. The previous argument shows that under $G_i$ the
representation $R$ breaks into a direct sum $R=r_1\oplus
r_2\oplus\cdots$ of at least two irreducible representations of
$G_i$. Therefore the product $R\times \bar R$, contains the sum of at
least two invariants under $G_i$, one in $r_1\times \bar r_1$ and
another in $r_2\times \bar r_2$.

\section{The Kaplan-Schmaltz model}
\label{sec:KS}
Famously, no-go theorems are most useful in showing how to avoid
them. There may be some  assumption one may be willing to give up. 
Kaplan and Schmaltz have studied a class of models for which our proof
fails.\cite{Kaplan:2003uc} Their models are peculiar in that
collective symmetries follow from setting the gauge coupling to zero
for some fields but not for others. That is where the model evades our
no-go argument. 

Our proof {\it assumes} that for each gauge group factor $G_i$ there
is one collective symmetry group $G^c_i$ that commutes with it. This
is useful because one can consider the limit in which all other gauge
couplings are set to zero and in that limit $G^c_i$ is an exact
symmetry. There is no such limit in Kaplan-Schmaltz (KM) models. In
them the collective symmetry limit is obtained by judiciously ignoring
certain terms in the Lagrangian, rather than by parametrically turning
them off.

Specifically, for Kaplan-Schmaltz there are two different collective
symmetry groups for the same gauge group factor. Neither of these
commutes with the gauge group, except if one ignores the gauge
couplings of a subset of fields.  For example, in their simplest model
$G_f=SU(3)_L\times SU(3)_R$, $H=SU(2)_L\times SU(2)_R$ and $G_w=
SU(3)_V$. As it stands there is no obvious collective symmetry. But
had we gauged $SU(3)_L$ only then $SU(3)_R$ would be a collective
symmetry group and vice versa. Since the order parameter is
$(3,1)+(1,3)$, one can accomplish this by ignoring the gauge coupling
of one or the other of $(3,1)$ or $(1,3)$. There is no fine tuning in
the Yukawa terms: collective symmetries are gauged $SU(3)$'s and
therefore the various EW invariants are now related. Of course, if the
symmetries are gauged then a higgs must be eaten. But since there are
two copies of $SU(3)$, there are two doublets, one is eaten and the
other is the higgs. This allows KS models to avoid the problem with
quartic couplings\cite{Schmaltz:2008vd} and, moreover, there is a
region of parameter space where it is consistent with electroweak
precision data\cite{Han:2005dz}.

\end{document}